# AN IMPLEMENTATION OF INTRUSION DETECTION SYSTEM USING GENETIC ALGORITHM

Mohammad Sazzadul Hoque[1], Md. Abdul Mukit[2] and Md. Abu Naser Bikas[3]

[1]Student, Department of Computer Science and Engineering, Shahjalal University of Science and Technology, Sylhet, Bangladesh
`sazzad@ymail.com`

[2]Student, Department of Computer Science and Engineering, Shahjalal University of Science and Technology, Sylhet, Bangladesh
`mukit.sust027@gmail.com`

[3]Lecturer, Department of Computer Science and Engineering, Shahjalal University of Science and Technology, Sylhet, Bangladesh
`bikasbd@yahoo.com`

## ABSTRACT

*Nowadays it is very important to maintain a high level security to ensure safe and trusted communication of information between various organizations. But secured data communication over internet and any other network is always under threat of intrusions and misuses. So Intrusion Detection Systems have become a needful component in terms of computer and network security. There are various approaches being utilized in intrusion detections, but unfortunately any of the systems so far is not completely flawless. So, the quest of betterment continues. In this progression, here we present an Intrusion Detection System (IDS), by applying genetic algorithm (GA) to efficiently detect various types of network intrusions. Parameters and evolution processes for GA are discussed in details and implemented. This approach uses evolution theory to information evolution in order to filter the traffic data and thus reduce the complexity. To implement and measure the performance of our system we used the KDD99 benchmark dataset and obtained reasonable detection rate.*

## KEYWORDS

*Computer & Network Security, Intrusion Detection, Intrusion Detection System, Genetic Algorithm, KDD Cup 1999 Dataset.*

## 1. INTRODUCTION

In 1987 Dorothy E. Denning proposed intrusion detection as is an approach to counter the computer and networking attacks and misuses [1]. Intrusion detection is implemented by an intrusion detection system and today there are many commercial intrusion detection systems available. In general, most of these commercial implementations are relative ineffective and insufficient, which gives rise to the need for research on more dynamic intrusion detection systems.

Generally an intruder is defined as a system, program or person who tries to and may become successful to break into an information system or perform an action not legally allowed [2]. We refer intrusion as any set of actions that attempt to compromise the integrity, confidentiality, or availability of a computer resource [3]. The act of detecting actions that attempt to compromise the integrity, confidentiality, or availability of a computer resource can be referred as intrusion detection [3]. An intrusion detection system is a device or software application that monitors network and/or system activities for malicious activities or policy violations and produces reports [4].

The remainder of the paper is organized as follows: Section 2 shortly describes some previous works. Section 3 gives an overview about intrusion detection system. Section 4 describes some existing intrusion detection systems and their problems. Section 5 and 6 describes our system and its implementation. Section 7 describes the performance analysis of our system. We conclude at section 8.

## 2. RELATED WORKS

Some import applications of soft computing techniques for Network Intrusion Detection is described in this section. Several Genetic Algorithms (GAs) and Genetic Programming (GP) has been used for detecting intrusion detection of different kinds in different scenarios. Some uses GA for deriving classification rules [5][6][7][8]. GAs used to select required features and to determine the optimal and minimal parameters of some core functions in which different AI methods were used to derive acquisition of rules [9][10][11]. There are several papers [12][13][14][15] related to IDS which has certain level of impact in network security.

The effort of using GAs for intrusion detection can be referred back to 1995, when Crosbie and Spafford [16] applied the multiple agent technology and GP to detect network anomalies [19]. For both agents they used GP to determine anomalous network behaviours and each agent can monitor one parameter of the network audit data. The proposed methodology has the advantage when many small autonomous agents are used but it has problem when communicating among the agents and also if the agents are not properly initialized the training process can be time consuming.

Li [6] described a method using GA to detect anomalous network intrusion [19][20]. The approach includes both quantitative and categorical features of network data for deriving classification rules. However, the inclusion of quantitative feature can increase detection rate but no experimental results are available.

Goyal and Kumar [18] described a GA based algorithm to classify all types of smurf attack using the training dataset with false positive rate is very low (at 0.2%) and detection rate is almost 100% [20].

Lu and Traore [7] used historical network dataset using GP to derive a set of classification [19]. They used support-confidence framework as the fitness function and accurately classified several network intrusions. But their use of genetic programming made the implementation procedure very difficult and also for training procedure more data and time is required

Xiao et al. [17] used GA to detect anomalous network behaviours based on information theory [19][20]. Some network features can be identified with network attacks based on mutual information between network features and type of intrusions and then using these features a linear structure rule and also a GA is derived. The approach of using mutual information and resulting linear rule seems very effective because of the reduced complexity and higher detection rate. The only problem is it considered only the discrete features.

Gong et al. [19] presented an implementation of GA based approach to Network Intrusion Detection using GA and showed software implementation. The approach derived a set of classification rules and utilizes a support-confidence framework to judge fitness function.

Abdullah et al. [20] showed a GA based performance evaluation algorithm to network intrusion detection. The approach uses information theory for filtering the traffic data.

# 3. INTRUSION DETECTION OVERVIEW

The below sections give a short overview of networking attacks, classifications and various components of Intrusion Detection System.

## 3.1. Networking Attacks

This section is an overview of the four major categories of networking attacks. Every attack on a network can comfortably be placed into one of these groupings [21].

- **Denial of Service (DoS):** A DoS attack is a type of attack in which the hacker makes a computing or memory resources too busy or too full to serve legitimate networking requests and hence denying users access to a machine e.g. apache, smurf, neptune, ping of death, back, mail bomb, UDP storm etc. are all DoS attacks.

- **Remote to User Attacks (R2L):** A remote to user attack is an attack in which a user sends packets to a machine over the internet, which s/he does not have access to in order to expose the machines vulnerabilities and exploit privileges which a local user would have on the computer e.g. xlock, guest, xnsnoop, phf, sendmail dictionary etc.

- **User to Root Attacks (U2R):** These attacks are exploitations in which the hacker starts off on the system with a normal user account and attempts to abuse vulnerabilities in the system in order to gain super user privileges e.g. perl, xterm.

- **Probing:** Probing is an attack in which the hacker scans a machine or a networking device in order to determine weaknesses or vulnerabilities that may later be exploited so as to compromise the system. This technique is commonly used in data mining e.g. saint, portsweep, mscan, nmap etc.

## 3.2. Classification of Intrusion Detection

Intrusions Detection can be classified into two main categories. They are as follow:

- **Host Based Intrusion Detection:** HIDSs evaluate information found on a single or multiple host systems, including contents of operating systems, system and application files [22].

- **Network Based Intrusion Detection:** NIDSs evaluate information captured from network communications, analyzing the stream of packets which travel across the network [22].

## 3.3. Components of Intrusion Detection System

An intrusion detection system normally consists of three functional components [23].

The first component of an intrusion detection system, also known as the event generator, is a **data source**. Data sources can be categorized into four categories namely Host-based monitors, Network-based monitors, Application-based monitors and Target-based monitors.

The second component of an intrusion detection system is known as the **analysis engine**. This component takes information from the data source and examines the data for symptoms of attacks or other policy violations. The analysis engine can use one or both of the following analysis approaches:

- **Misuse/Signature-Based Detection:** This type of detection engine detects intrusions that follow well-known patterns of attacks (or signatures) that exploit known software vulnerabilities [24][25]. The main limitation of this approach is that it only looks for the known weaknesses and may not care about detecting unknown future intrusions [26].

▪ **Anomaly/Statistical Detection:** An anomaly based detection engine will search for something rare or unusual [26]. They analyses system event streams, using statistical techniques to find patterns of activity that appear to be abnormal. The primary disadvantages of this system are that they are highly expensive and they can recognize an intrusive behavior as normal behavior because of insufficient data

▪ The third component of an intrusion detection system is the ***response manager***. In basic terms, the response manager will only act when inaccuracies (possible intrusion attacks) are found on the system, by informing someone or something in the form of a response.

## 4. EXISTING SYSTEMS AND THEIR PROBLEMS

Here we describe some of the important Intrusion Detection systems and their problems.

### 4.1. Existing Intrusion Detection Systems

▪ **Snort:** A free and open source network intrusion detection and prevention system, was created by Martin Roesch in 1998 and now developed by Sourcefire. In 2009, Snort entered InfoWorld's Open Source Hall of Fame as one of the "greatest open source software of all time" [36][37]. Through protocol analysis, content searching, and various pre-processors, Snort detects thousands of worms, vulnerability exploit attempts, port scans, and other suspicious behavior [34][35].

▪ **OSSEC:** An open source host-based intrusion detection system, performs log analysis, integrity checking, rootkit detection, time-based alerting and active response [34][35]. In addition to its IDS functionality, it is commonly used as a SEM/SIM solution. Because of its powerful log analysis engine, ISPs, universities and data centers are running OSSEC HIDS to monitor and analyze their firewalls, IDSs, web servers and authentication logs.

▪ **OSSIM:** The goal of Open Source Security Information Management, OSSIM is to provide a comprehensive compilation of tools which, when working together, grant network/security administrators with a detailed view over each and every aspect of networks, hosts, physical access devices, and servers [35]. OSSIM incorporates several other tools, including Nagios and OSSEC HIDS.

▪ **Suricata:** An open source-based intrusion detection system, was developed by the Open Information Security Foundation (OISF) [38].

▪ **Bro**: An open-source, Unix-based network intrusion detection system [39]. Bro detects intrusions by first parsing network traffic to extract its application-level semantics and then executing event-oriented analyzers that compare the activity with patterns deemed troublesome.

▪ **Fragroute/Fragrouter:** A network intrusion detection evasion toolkit [34]. Fragrouter helps an attacker launch IP-based attacks while avoiding detection. It is part of the NIDSbench suite of tools by Dug Song.

▪ **BASE:** The Basic Analysis and Security Engine, BASE is a PHP-based analysis engine to search and process a database of security events generated by various IDSs, firewalls and network monitoring tools [34].

▪ **Sguil:** Sguil is built by network security analysts for network security analysts [34][35]. Its main component is an intuitive GUI that provides real-time events from Snort/barnyard. It also includes other components which facilitate the practice of network security monitoring and event driven analysis of IDS alerts.

## 4.2. Problems with Existing Systems

Most existing intrusion detection systems suffer from at least two of the following problems [2]:

- First, the information used by the intrusion detection system is obtained from audit trails or from packets on a network. Data has to traverse a longer path from its origin to the IDS and in the process can potentially be destroyed or modified by an attacker. Furthermore, the intrusion detection system has to infer the behavior of the system from the data collected, which can result in misinterpretations or missed events. This is referred as the *fidelity* problem.

- Second, the intrusion detection system continuously uses additional resources in the system it is monitoring even when there are no intrusions occurring, because the components of the intrusion detection system have to be running all the time. This is the *resource usage* problem.

- Third, because the components of the intrusion detection system are implemented as separate programs, they are susceptible to tampering. An intruder can potentially disable or modify the programs running on a system, rendering the intrusion detection system useless or unreliable. This is the *reliability* problem.

## 5. Our IDS using Genetic Algorithm

We have chosen GA to make our intrusion detection system. This section gives an overview of the algorithm and the system.

### 5.1. Genetic Algorithm Overview

A Genetic Algorithm (GA) is a programming technique that mimics biological evolution as a problem-solving strategy [27]. It is based on Darwinian's principle of evolution and survival of fittest to optimize a population of candidate solutions towards a predefined fitness [6].

GA uses an evolution and natural selection that uses a chromosome-like data structure and evolve the chromosomes using selection, recombination and mutation operators [6]. The process usually begins with randomly generated population of chromosomes, which represent all possible solution of a problem that are considered candidate solutions. From each chromosome different positions are encoded as bits, characters or numbers. These positions could be referred to as genes. An evaluation function is used to calculate the goodness of each chromosome according to the desired solution; this function is known as "Fitness Function". During the process of evaluation "Crossover" is used to simulate natural reproduction and "Mutation" is used to mutation of species [6]. For survival and combination the selection of chromosomes is biased towards the fittest chromosomes.

When we use GA for solving various problems three factors will have vital impact on the effectiveness of the algorithm and also of the applications [19]. They are: i) the fitness function; ii) the representation of individuals; and iii) the GA parameters. The determination of these factors often depends on applications and/or implementation.

## 5.2. Flowchart

Figure 1 shows the operations of a general genetic algorithm according to which GA is implemented into our system.

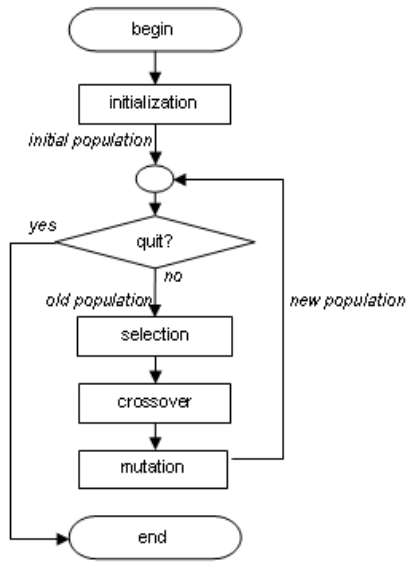

Figure 1.  Flowchart of Genetic Algorithm

## 5.3. Algorithm of Our System

Our system can be divided into two main phases: the precalculation phase and the detection phase. Listing 1 depicts major steps in precalculation phase, where a set of chromosome is created using training data. This chromosome set will be used in the next phase for the purpose of comparison.

Listing 1.  Major steps in precalculation

```
Algorithm : Initialize chromosomes for comparison
Input : Network audit data (for training)
Output : A set of chromosomes

1. Range = 0.125
2. For each training data
3. If it has neighboring chromosome within Range
4. Merge it with the nearest chromosome
5. Else
6. Create new chromosome with it
7. End if
8. End for
```

Listing 2 depicts major steps of detection phase, where a population is being created for a test data and going through some evaluation processes (selection, crossover, mutation) the type of the test data is predicted. The precalculated set of chromosome is used in this phase to find out fitness of each chromosome of the population.

Listing 2.  Major steps in detection

```
Algorithm : Predict data/intrusion type (using GA)
Input : Network audit data (for testing), Precalculated set of chromosomes
Output : Type of data.

1. Initialize the population
2. CrossoverRate = 0.15, MutationRate = 0.35
3. While number of generation is not reached
4. For each chromosome in the population
5. For each precalculated chromosome
6. Find fitness
7. End for
8. Assign optimal fitness as the fitness of that chromosome
9. End for
10. Remove some chromosomes with worse fitness
11. Apply crossover to the selected pair of chromosomes of the population
12. Apply mutation to each chromosome of the population
13. End while
```

## 6. OUR IMPLEMENTATION

To implement our algorithm and to evaluate the performance of our system, we have used the standard dataset used in KDD Cup 1999 "Computer network intrusion detection" competition.

### 6.1. KDD Sample Dataset

For the implementation of our algorithm we used the KDD 99 intrusion detection datasets which are based on the 1998 DARPA initiative, which provides designers of intrusion detection systems (IDS) with a benchmark on which to evaluate different methodologies [28][32]. Hence, a simulation is being made of a factitious military network with three 'target' machines running various operating systems and services. They also used three additional machines to spoof different IP addresses for generate network traffic.

A connection is a sequence of TCP packets starting and ending at some well defined times, between which data  flows from a source IP address to a target IP address under some well defined protocol [28][29][32]. It results in 41 features for each connection.

Finally, there is a sniffer that records all network traffic using the TCP dump format [32]. The total simulated period is seven weeks. Normal connections are created to profile that expected in a military network and attacks fall into one of four categories: User to Root; Remote to Local; Denial of Service; and Probe.

The KDD 99 intrusion detection benchmark consists different components [30]:
kddcup.data;        kddcup.data_10_percent;        kddcup.newtestdata_10_percent_unlabeled;
kddcup.testdata.unlabeled; kddcup.testdata.unlabeled_10_percent; corrected.

We have used "kddcup.data_10_percent" as training dataset and "corrected" as testing dataset. In this case the training set consists of 494,021 records among which 97,280 are normal connection records, while the test set contains 311,029 records among which 60,593 are normal connection records. Table 1 shows the distribution of each intrusion type in the training and the test set.

Table 1.  Distribution of intrusion types in datasets

| Dataset | normal | probe | dos | u2r | r2l | Total |
|---|---|---|---|---|---|---|
| Train ("kddcup.data_10_percent") | 97280 | 4107 | 391458 | 52 | 1124 | 494021 |
| Test ("corrected") | 60593 | 4166 | 229853 | 228 | 16189 | 311029 |

## 6.2. Implementation Procedure

In the precalculation phase, we have made 23 groups of chromosomes according to training data. There were 23 (22+1) groups for each of attack and normal types presented in training data. Number of chromosomes in each group is variable and depends on the number of data and relationship among data in that group. Total number of chromosomes in all groups were tried to keep in reasonable level to optimize time consumption in testing phase.

In the testing / detection phase, for each test data, an initial population is made using the data and occurring mutation in different features. This population is compared with each chromosomes prepared in training phase. Portion of population, which are more loosely related with all training data than others, are removed. Crossover and mutation occurs in rest of the population which becomes the population of new generation. The process runs until the generation size comes down to 1 (one). The group of the chromosome which is closest relative of only surviving chromosome of test data is returned as the predicted type.

Among the extracted features of the datasets, we have taken only the numerical features, both continuous and discrete, under consideration for the sake of the simplification of the implementation.

## 7. EXPERIMENTAL RESULTS AND ANALYSIS

From our system we get the confusion matrics depicted in table 2. For most of the classes, our system performs well enough except normal data type which is because of ignoring non-numerical features. Comparing with the confusion matrics of the winning entry of KDD'99 [31], we get better detection rate for denial of service & user-to-root and close detection rate for probe & remote-to-local.

Table 2.  Confusion metrics for system evaluation

| | | Predicted label | | | | | % Correct |
|---|---|---|---|---|---|---|---|
| | | normal | probe | dos | u2r | r2l | |
| Actual class | normal | 42138 | 1421 | 15835 | 486 | 713 | 69.5% |
| | probe | 398 | 2963 | 654 | 2 | 149 | 71.1% |
| | dos | 921 | 432 | 228489 | 1 | 10 | 99.4% |
| | u2r | 146 | 21 | 8 | 43 | 10 | 18.9% |
| | r2l | 11191 | 578 | 3398 | 141 | 881 | 5.4% |
| % Correct | | 76.9% | 54.7% | 92.0% | 6.4% | 50.0% | |

For simplified evaluation of our system, besides the classical accuracy measure, we have used two standard metrics of detection rate and false positive rate developed for network intrusions derived in [33]. Table 3 shows these standard metrics.

Table 3. Standard metrics for system evaluation

| | | Predicted label | |
|---|---|---|---|
| | | Normal | Intrusion |
| Actual Class | Normal | True Negative (42138) | False Positive (18455) |
| | Intrusion | False Negative (12528) | True Positive (237908) |

Detection rate for each data type can be seen from figure 2.

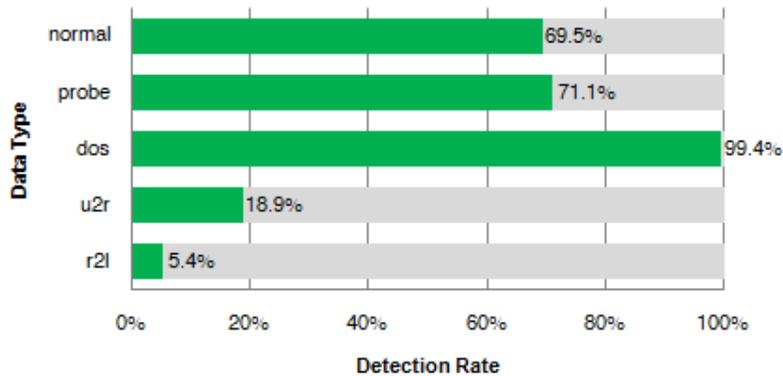

Figure 2. Detection rate for each class

Detection rate (DR) is calculated as the ratio between the number of correctly detected intrusions and the total number of intrusions [33], that is:

$$DR = \frac{\#TruePositive}{\#FalseNegative + \#TruePositive}$$

Using table 3, detection rate, DR = 0.9500.

False positive rate (FP) is calculated as the ratio between the numbers of normal connections that are incorrectly classifies as intrusions and the total number of normal connections [33], that is:

$$FP = \frac{\#FalsePositive}{\#TrueNegative + \#FalsePositive}$$

Using table 3, false positive rate, FP = 0.3046.

## 8. CONCLUSIONS

In this paper, we present and implemented an Intrusion Detection System by applying genetic algorithm to efficiently detect various types of network intrusions. To implement and measure the performance of our system we used the standard KDD99 benchmark dataset and obtained reasonable detection rate. To measure the fitness of a chromosome we used the standard deviation equation with distance. If we can use a better equation or heuristic in this detection process we believe the detection rate and process will improve a great extent, especially false positive rate will surely be much lower. In near future we will try to improve our intrusion detection system with the help of more statistical analysis and with better and may be more complex equations.

## Authors

**Mohammad Sazzadul Hoque**

Mohammad Sazzadul Hoque is a B.Sc. student in the Dept. of Computer Science and Engineering, Shahjalal University of Science and Technology, Bangladesh. His research interest includes Computer & Network Security, Intrusion Detection and Intrusion Prevention.

**Md. Abdul Mukit**

Md. Abdul Mukit is a B.Sc. student in the Dept. of Computer Science and Engineering, Shahjalal University of Science and Technology, Bangladesh. His research interest includes Computer & Network Security, Intrusion Detection and Intrusion Prevention.

**Md. Abu Naser Bikas**

Md. Abu Naser Bikas obtained his B. Sc. Degree in Computer Science & Engineering from Shahjalal University of Science & Technology, Bangladesh. Currently, he is a Lecturer in Computer Science & Engineering department at the same University. His research interests include Network Security, Intrusion Detection and Intrusion Prevention, VANET, Bangla OCR and Grid Computing. He has published approximately 12 research papers in reputed International journals and proceedings.